\documentclass[wsdraft]{ws-procs11x85}
\usepackage{ws-procs-thm}           

\usepackage[final]{fixme}
\fxsetup{inline,nomargin,theme=color}

\date{}
\begin{document}

\title{Robustly Extracting Medical Knowledge from EHRs: \\ A Case Study of Learning a Health Knowledge Graph
}

\author{Irene Y. Chen$^\dag$, Monica Agrawal$^\dag$, Steven Horng$^*$, and David Sontag$^\dag$}

\address{$^\dag$Electrical Engineering and Computer Science, Massachusetts Institute of Technology,\\
Cambridge, MA, USA\\ $^*$Department of Emergency Medicine, Beth Israel Deaconess Medical Center,\\
Boston, MA, USA\\
Corresponding e-mail: iychen@mit.edu\\}



\begin{abstract}
Increasingly large electronic health records (EHRs) provide an opportunity to algorithmically learn medical knowledge. In one prominent example, a causal health knowledge graph could learn relationships between diseases and symptoms and then serve as a diagnostic tool to be refined with additional clinical input. Prior research has demonstrated the ability to construct such a graph from over 270,000 emergency department patient visits. In this work, we describe methods to evaluate a health knowledge graph for robustness. Moving beyond precision and recall, we analyze for which diseases and for which patients the graph is most accurate. We identify sample size and unmeasured confounders as major sources of error in the health knowledge graph. We introduce a method to leverage non-linear functions in building the causal graph to better understand existing model assumptions. Finally, to assess model generalizability, we extend to a larger set of complete patient visits within a hospital system. We conclude with a discussion on how to robustly extract medical knowledge from EHRs. 
\end{abstract}

\keywords{knowledge graph, healthcare, electronic health records, machine learning}

\copyrightinfo{\copyright\ 2019 The Authors. Open Access chapter published by World Scientific Publishing Company and distributed under the terms of the Creative Commons Attribution Non-Commercial (CC BY-NC) 4.0 License.}

\section{Introduction}
\label{sec:intro}
Clinicians are often interested in the causal relationship between diseases and symptoms. Given presenting symptoms, what is the likely diagnosis? An accurate understanding of these relations can be leveraged to build diagnostic tools, which have been shown to be useful for training clinicians,~\cite{miller1989use} assisting clinical workflows,~\cite{barnett1987dxplain} or even in substitution of clinicians.~\cite{de1972computer} 

Historically, researchers have relied on manual creation of such diagnostic tools, which can be expensive and brittle. One of the first attempts, the Quick Medical Reference (QMR) tool, sought to create ``on the fly" medical information to suit a clinician's queries and famously took hundreds of thousands of clinician-hours to assemble.~\cite{shwe1991probabilistic} Successors to QMR faced similar challenges because of the same reliance on expert information.~\cite{miller1994medical} Adapting a manually created diagnostic tool to new diseases and new symptoms relies on enormous human capacity. Any differences in patient populations present another obstacle to the usefulness of manually created models. Lastly, physicians are exceptionally poor at accurately approximating the probability of rare diseases and symptoms. For example, physicians can yield to availability bias, inadvertently overestimating the probability of a rare disease they recently read about, seen, or are emotionally linked to. Physicians often also approximate rare symptoms with a probability of 0, saying that they are impossible rather than unlikely.

The increased availability of electronic health knowledge records allows researchers to readily learn latent patterns from observational data. In contrast to clinical trials which are conducted on restricted subpopulations,~\cite{antman1985selection,cottin1999small,masoudi2003most,herland2005representative} the breadth of electronic health records (EHRs) allows for the inclusion of all patients who enter the healthcare system.~\cite{jensen2012mining} With this data, researchers can build models to extract general medical knowledge~\cite{finlayson2014building,rotmensch2017learning,sondhi2012sympgraph} and diagnose patients~\cite{fatima2017survey, gargeya2017automated,hannun2019cardiologist}.

Although diagnostic knowledge exists in medical textbooks~\cite{mcphee2010current} or online repositories like Mayo Clinic~\cite{hypothyroidism}, inferring that medical knowledge automatically from EHRs provides different strengths. First, because EHRs provide a dramatically different perspective than idealized and curated medical textbooks, we may find new connections between diseases and symptoms. Second, the automated nature of extraction allows for any large EHR system to be used as source dataset. Armed with medical knowledge gleaned from EHRs and canonical medical training, clinicians can leverage both for improved clinical decision making.

However, the use of EHRs to build health knowledge graphs is not without potential bias. Because of the nature of data collection, models learned on EHRs like health knowledge graphs are subject to many sources of statistical bias~\cite{gianfrancesco2018potential,ghassemi2018opportunities}. First, narrow sample sizes of subsets of the data can cause underfitting despite the larger scale of the entire dataset. Second, confounders not measured by the data may bias the reliability of resulting models.~\cite{pearl2000causality} Lastly, algorithms or findings from algorithms may not generalize to entirely different populations. 

It is essential that we closely examine so-called health knowledge graphs so that they can be used as the first steps of a diagnostic tool to improve clinical workflow and better understand diseases. We seek to understand for which diseases and for which patients a health knowledge graph performs poorly and understand potential confounders. Error analysis can guide steps to improve derived models. For example, a change in model formulation could improve performance if sources of error are well understood. Moreover, data augmentation through additional features or a more broadly collected dataset could provide additional signal to the health knowledge graph. We hope that this critical evaluation of extracted health knowledge graphs provides an example on how to assess the robustness of medical knowledge extracted from EHRs. 

\paragraph{Contributions.}
We present a methodology for analyzing a graph relating diseases and symptoms derived from EHRs. We explicitly address open questions raised in prior work~\cite{rotmensch2017learning} in an acute care setting from over 270,000 patient records. Open questions include error analysis on the learned health knowledge graph and the inclusion of non-linear models. Similar to prior work~\cite{rotmensch2017learning}, we use a manually curated health knowledge graph from Google for an automated evaluation.~\cite{ramaswami_2015} We aim to provide guidance to researchers about the relationship between dataset size, model specification, and performance for constructing health knowledge graphs.

We describe our main findings: 
\begin{enumerate}
    
\item We identify and analyze diseases with the lowest precision and recall in existing health knowledge graphs. In particular, diseases with lower performance correlate with having more co-occurring diseases, more co-occurring symptoms, and fewer observations. 
\item We present results on the impact of adding demographic data of age and gender. We find that demographic data improves performance for certain diseases and therefore some models. We find that differences in graphs learned from patients of varying age groups and different genders can be explained by drastically smaller sample sizes and differences in presentations. 

\item We propose a method for learning a health knowledge graph from non-linear models based on principles from causal inference. We show that these causal methods do not yield an increase in performance as the noisy OR approach is surprisingly robust in its performance in a wide range of datasets and settings.

\item We compare the health knowledge graph learned in the emergency department setting to ones learned from the complete online medical records of the same patients. We find that differences in coarseness yield a change in ranking of performance between models, and we propose potential sources of confounding.

\end{enumerate}

We conclude with a discussion on robustly extraction of medical knowledge from EHRs. 

\section{Related Work}
\label{sec:related}

In the most directly related work, researchers demonstrated how to learn a health knowledge graph from over 270,000 patient visits to the emergency department at Beth Israel Deaconess Medical Center~\cite{rotmensch2017learning}. Health knowledge graphs were extracted from the clinical records and maximum likelihood estimation of three models: logistic regression, naive Bayes, and noisy OR. 

We formalize a graph of health knowledge with diseases and symptoms as nodes. When a disease (e.g. type 1 diabetes) causes a symptom (e.g. excessive thirst), an edge is drawn between corresponding nodes. For each model, an importance metric is constructed to determine whether an edge should exist between a disease and symptom. For logistic regression, the importance metric $\delta_{ij}$ is $\delta_{ij} = \max (0, b_{ij})$ where $b_{ij}$ is the weight associated with symptom $i$ in a logistic regression fit to predict disease $j$. For naive Bayes, the importance metric is 
$$\delta_{ij} = \log(p(X_i = 1 | Y_j = 1)) - \log(p(X_i = 1 | Y_j = 0))$$
where $X_i$ is the binary variable denoting the presence of symptom $i$ and $Y_j$ is the binary variable indicating the presence of disease $j$. Lastly, for noisy OR, we denote importance metric $f_{ij}$ as the probability that a disease $Y_j$ of $D$ total diseases will turn on symptom $x_i$ while taking into account leak probability $l_i$ that turns on symptom $i$ regardless. The probability of symptom $X_i$ being observed is
$$p(X_i = 1 | Y_1, \ldots, Y_D) = 1 - (1-l_i) \prod_j (f_{ij})^{Y_j}$$
The corresponding importance metric for noisy OR is $\delta_{ij} = 1 - f_{ij}$. For each model, a threshold is chosen, and disease-symptom pairs with an importance metric over the threshold are included as an edge in a bipartite graph of diseases and symptoms. 

We note that of the three methods, logistic regression and naive Bayes predict disease occurrence from observed symptoms whereas noisy OR predicts the probability of a symptom occurrence given the observed diseases. Naive Bayes and noisy OR can be reformulated to show the causal relationship between diseases and symptoms whereas logistic regression does not have an explicit tie to causal inference. We introduce a method to use logistic regression and non-linear methods to build an explicitly causal relationship in Section~\ref{sec:nonlinear}.

The derived graphs were evaluated against expert physician opinion and, with permission, a manually curated health knowledge graph from Google. The health knowledge graph created from parameters of a fitted noisy OR model had the best precision and recall, with prior results reported in the first columns of Table~\ref{tab:demo}.

Other similar works have examined the associative rather than causal link between diseases and symptoms in the clinical records~\cite{finlayson2014building} or estimating the strength of an edge between diseases and symptoms by the distance of mentions in the clinical records.~\cite{sondhi2012sympgraph}

\section{Methods}
\label{sec:methods}
\subsection{Data collection and preparation}
\label{sec:data}
We consider two datasets; both were collected from patient visits to Beth Israel Deaconess Medical Center (BIDMC), a trauma center and tertiary academic teaching hospital with approximately 55,000 visits/year to the emergency department (ED). The first is the original ED dataset of 273,174 patient de-identified records. It includes all 140,804 patients who visited the ED between 2008 and 2013. The second is a new and related dataset consisting of the complete records (CR) of 140,446 of the patients from the ED dataset. The CR dataset includes 7,401,040 notes, spanning additional sources of data (e.g. cardiology or primary care physician notes).

For the ED dataset, each record represents a single patient visit. For the CR dataset, we consider all notes for a patient (a median of 13 notes each), which at the median, span approximately 2.5 years. We create three separate datasets from the CR dataset, each using a different way of aggregating a patient's timeline. First, we treat every note in a patient's file as a separate instance, yielding 7,401,040 instances. Second, in order to capture associations across temporally close notes, but avoid spurious associations from unrelated notes, we split each patient's file into a variable number of episodes. We defined an episode as a sequence of notes in which any two consecutive notes are at most 30 days apart. If a patient has a gap in her record of more than 30 days, the next record will create a new episode. This process yields 1,481,283 patient episodes, with a median of 3 episodes per patient and 2 notes per episode. Lastly, we consider an entire patient record for each patient as a single instance, ignoring the time component of the medical record. This decision yields 140,446 patient instances, corresponding to the number of unique patients in the dataset. 

We extracted positive mentions of 192 possible diseases and 771 possible symptoms from both datasets. For each ED visit or CR episode, the mentions are stored as binary variables to indicate the observation of a disease or symptom. For the ED dataset, concepts were extracted from structured data (e.g. ICD-9 diagnosis codes) as well as unstructured data (e.g. chief complaint, triage assessments, doctor's comments). For the CR dataset, we extract mentions only from unstructured notes. Unstructured data extraction was conducted using string matching.

We defined sufficient support for a disease as having at least 100 positive mentions and for a symptom as having at least 10 positive mentions. For the ED dataset, these constraints resulted in 156 diseases and 491 symptoms, and we extracted the same diseases and symptoms in the CR dataset. Age and gender variables are extracted from the patient records at the time of ED admission.

\subsection{Evaluation with GHKG}

Our source of automated evaluation comes from a health knowledge graph manually curated by Google.~\cite{ramaswami_2015} The Google Health Knowledge Graph (GHKG) represents a combined effort of several physicians through a multi-step process in a precise but not necessarily complete graph. The GHKG has been shown to be consistent with manually-compiled expert clinician opinion.~\cite{rotmensch2017learning}

To use the GHKG for evaluation, we compare the existence of a disease-symptom edge in a proposed health knowledge graph and the same disease-syptom edge in the GHKG. We use the resulting precision and recall as evaluation metrics. For individual diseases, we report the F1 score.~\cite{van1980new} For assessing health knowledge graphs recovered by likelihood estimation, we use the area under the precision-recall curve (AUPRC). For each disease, we include the same number of edges as in the GHKG. For additional details in implementing the F1 score and the AUPRC, see the supplementary materials. “Pain” is removed from the available symptoms because of inconsistent inclusion in the GHKG.

\subsection{Disease predictability analysis}
\label{sec:disease_analysis}

To compare performance for individual diseases, for each extracted disease, we assess the existence of an edge between that disease and all possible symptoms for a proposed model and compare against the GHKG to compute the F1 score as a function of precision and recall. For each disease, we sort symptoms by their importance scores and include the top 25 symptoms in the health knowledge graph as a binary edge between disease and symptom. Over all patients with each disease, we compute the number of occurrences of the disease, average number of extracted diseases per patient, average number of extracted symptoms per patient, average patient age, and percentage female. 

We denote a disease as abnormal --- with a corresponding column label in Table~\ref{tab:top_bottom} --- if the disease has a:
\begin{itemize}
    \item number of occurrences a standard deviation below the average number of occurrences (\texttt{count})
    \item mean number of extracted diseases a standard deviation above the mean extracted diseases (\texttt{disease})
    \item mean number of extracted symptoms a standard deviation above the mean extracted symptoms (\texttt{symptom})
    \item mean patient age a standard deviation either above or below the population mean (\texttt{age})
    \item a female percentage a standard deviation either above or below the population female percentage (\texttt{female})
    \item any of the above abnormalities (\texttt{any})
\end{itemize}

For a more detailed description of the disease error analysis, please see the supplementary materials. We sort the diseases in order of F1 scores and compare the 50 diseases with the highest F1 scores with the 50 diseases with the lowest F1 scores. When the standard deviation exceeds the range of the value, we use a percentile approximation.

\subsection{Demographic analysis}
\label{sec:methods_demo}

We include the demographic information in the health knowledge graph learning. We use the age and sex recorded at the time of the emergency room visit for each patient. We refer the reader to the supplementary materials for additional information about demographic data extraction and graphs of age and sex distribution in the dataset. 

\subsection{Non-linear methods}
\label{sec:nonlinear}
We propose a new method of introducing non-linear functions to estimate the existence of a disease-symptom edge. Leveraging notions from causal inference, we want to designate an edge from disease to symptom if the probability of observing a symptom is higher when we force a disease to occur than when we force a disease to not occur. Using the notion of a do-operator,~\cite{pearl2000causality} we formalize an importance measure $\delta_{ij}$ between symptom $X_i$ and condition $Y_j$
$$\delta_{ij} = \frac{\mathbb{P}(X_i = 1 | do(Y_j = 1))}{\mathbb{P}(X_i = 1| do(Y_j = 0))}$$

Using our EHRs, we can estimate $\mathbb{P}(X_i = x_i | do(Y_j = y_j))$ by considering each $x_{ni}, y_{nj}$ as the binary features for whether patient stay $n$ has symptom $i$ or condition $j$.
$$\mathbb{P}(X_i = 1 | do(Y_j = y_j)) \approx \frac{1}{N} \sum_{n=1}^N \hat{\mathbb{P}} (x_{ni} | do(Y_{nj} = y_j))$$
where $do(Y_j = 1)$ indicates intervening on diseases $Y_j$.

Using a calibrated and accurate predictive model (e.g. random forest, logistic regression) for $\hat{P}$, we can compute $\hat{\mathbb{P}} (x_{ni} | do(y_{nj}))$ --- where $y_{nj}$ indicates whether disease $j$ was observed at patient visit $n$ --- through fitting a function on the predicted probability for each patient stay and intervening to set $Y_{nj} = y_{nj}$. Our method allows the introduction of non-linear functions although we can and do use linear functions as $\hat{P}$.

Note that prior work~\cite{rotmensch2017learning} using logistic regression and naive Bayes relies on predicting disease $Y_j$ from the observation of symptoms $X_1, \ldots, X_S$ for $S$ symptoms. Similar to the noisy OR, we focus on predicting a symptom $X_i$ from observed diseases $Y_1, \ldots, Y_D$ for $D$ symptoms.

We use a logistic regression, random forest, and naive Bayes as our potential predictive model and denote the corresponding health knowledge graphs as ``causal" in Table~\ref{tab:demo} to avoid confusion with the prior logistic regression method. For each symptom learned, we conduct a hyperparameter search using 3-fold cross validation. The final health knowledge graph results from the model trained using the parameters that yielded the best cross-validation area under the receiver operator curve (AUROC). For logistic regression, we search over penalty norm (L1, L2) and 10 regularization parameters between 0.001 and 10, evenly spaced logarithmically. For random forest, we search over 8 maximum depths between 2 and 1024 and 8 values of minimum samples per leaf between 10 and 200, evenly spaced logarithmically. 



\section{Results}
\label{sec:results}

\begin{figure}
    \centering
    \includegraphics[width=0.4 \textwidth]{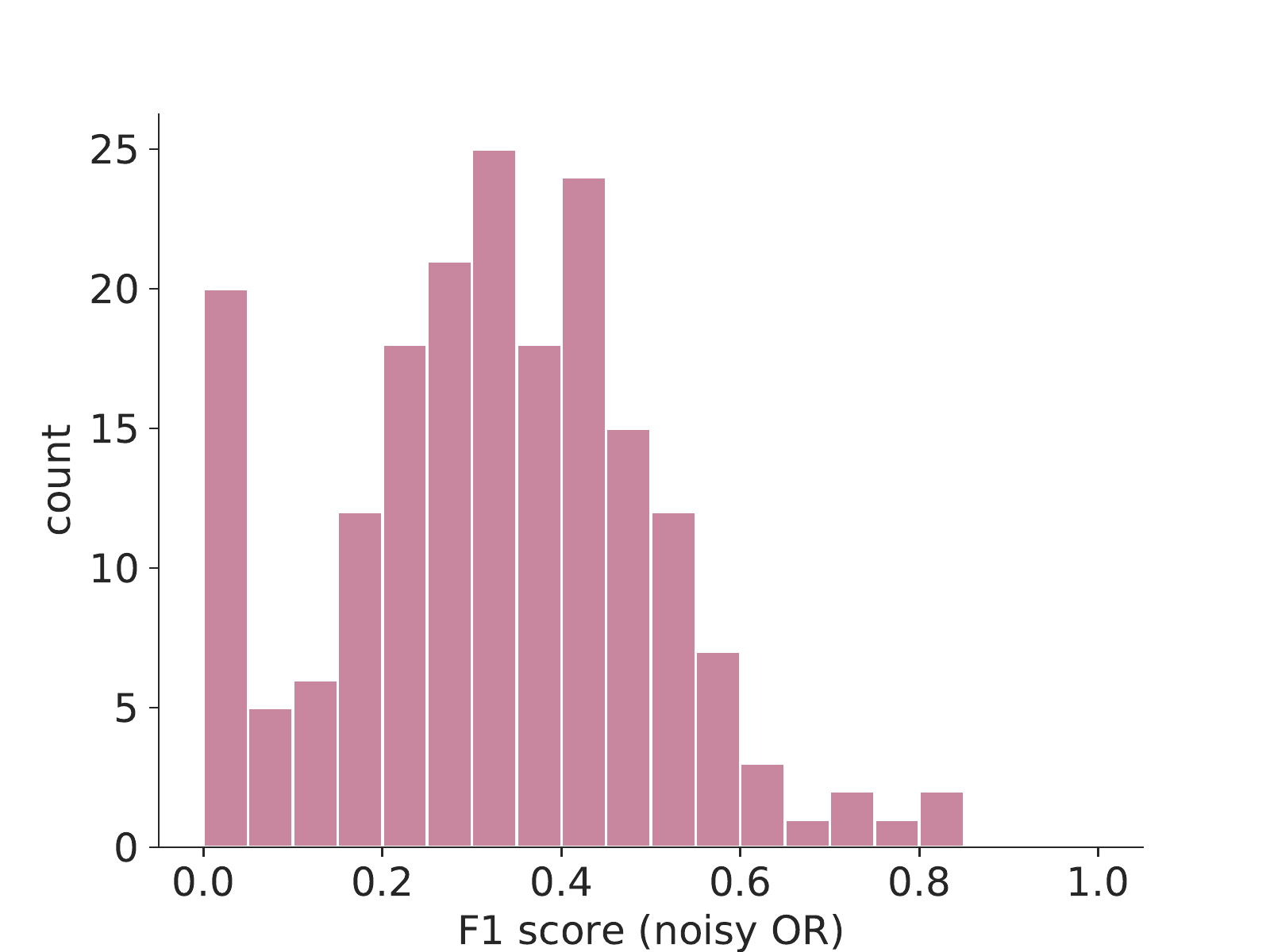}
    \includegraphics[width=0.4\textwidth]{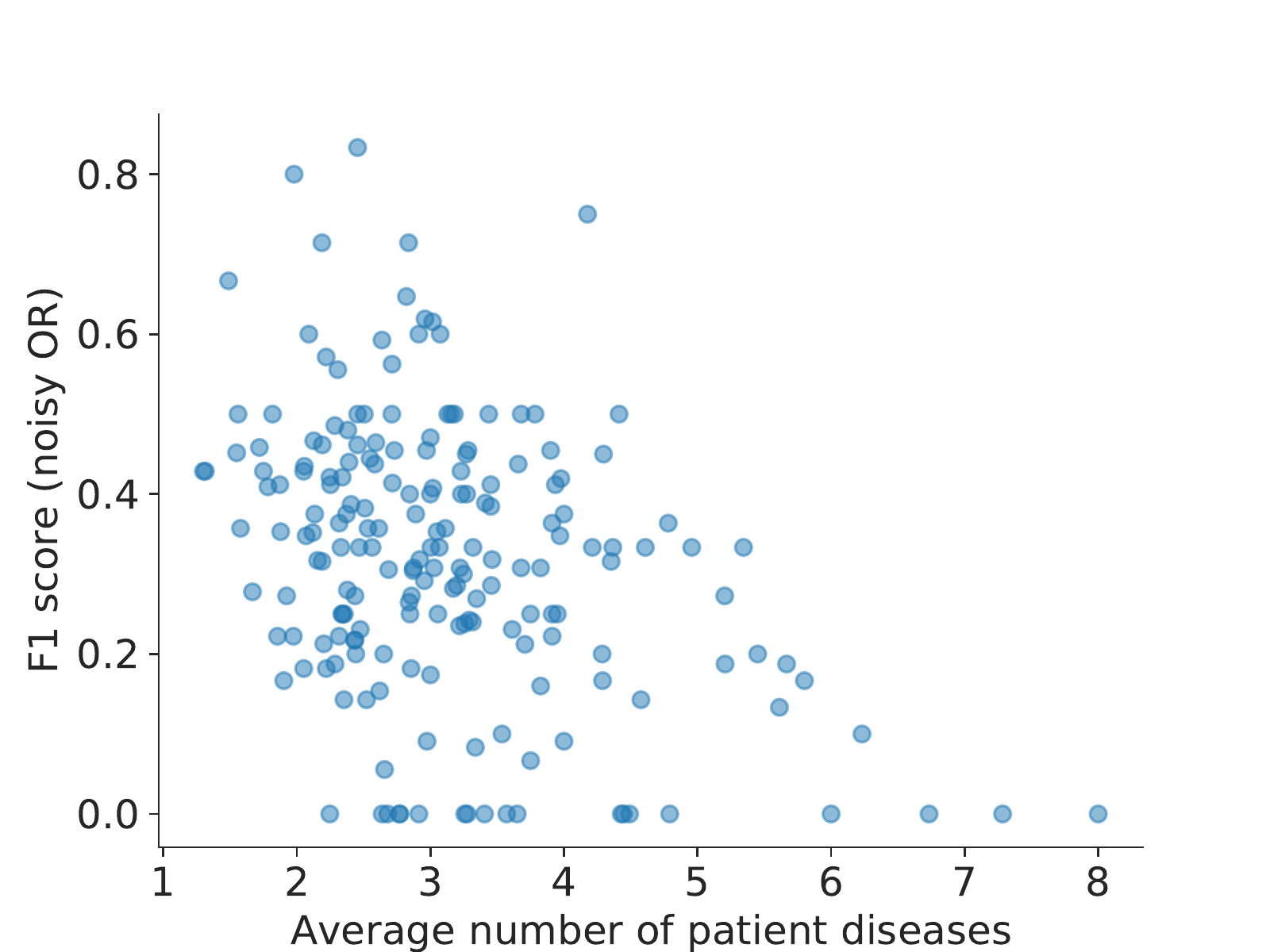}
    \caption{\textbf{Left:} Distribution of F1 scores by disease for noisy OR on the ED dataset. \textbf{Right:} F1 score for noisy OR model decreases with number of diseases per patient.}
    \label{fig:diseases}
\end{figure}

We first assess the diseases for which noisy OR, the highest performing health knowledge graph~\cite{rotmensch2017learning}, has the lowest and highest performance on the ED dataset. In Figure~\ref{fig:diseases} (left), we see that the distribution of F1 scores of diseases in the noisy OR health knowledge graph span high and low values. In Figure~\ref{fig:diseases} (right), the F1 scores decrease for diseases that have patients with many co-occurring diseases, which may indicate that the multiple diseases are reducing the signal that noisy OR can learn. 

We sort diseases by F1 score in the noisy OR health knowledge graph in Table~\ref{tab:top_bottom}. As defined in Section~\ref{sec:disease_analysis}, abnormalities occur at much higher rates for diseases with lower F1 scores. Compared to diseases with higher F1 scores, diseases with lower F1 scores are more likely to have more co-occurring diseases, more co-occurring symptoms, more extremely low or extremely high ages, and more extremely low or extremely high female percentages. Notably the number of occurrences does not appear to be associated with high or low F1 scores, probably due to careful selection of diseases and symptoms in the problem construction. For example, obesity has an extremely low F1 score of 0.11, which may be explained by high number of diseases among patients with obesity of 4.3 (compared to the mean of 3.0). In total, 66\% of diseases with the lowest 50 F1 scores have some defined aberration compared to 40\% of diseases with the highest 50 F1 scores.

\begin{table}[h]
    \centering
     \tbl{Percentage of diseases with abnormalities, as outlined in Section~\ref{sec:disease_analysis}, sorted by noisy OR F1 score, learned on ED data.}
    {
    \begin{tabular}{l  c c c c c  c}
         & \texttt{count} & \texttt{disease} & \texttt{symptom} & \texttt{age} & \texttt{female} & \texttt{any}  \\
        \hline
        \hline 
        top 50 & 16\% & 18\% & 8\% & 10\% & 8\% & 40\% \\
        bottom 50 & 16\% & 32\% & 16\% & 16\% & 24\% & 66\% \\

    \end{tabular}
   
    \label{tab:top_bottom}
    }
\end{table}

The abnormalities in age and gender suggest that including demographic information like age and gender may improve existing models. In Table~\ref{tab:demo}, we find that the addition of age and sex does not meaningful improve existing models. For space, we abbreviate logistic regression (LR), naive Bayes (NB), random forest (RF), and noisy OR (NO). Models using demographic information are designated with ``demo", models using the do-operator formulation are denoted ``causal".

\begin{table}[h]
\begin{minipage}{3in}

\centering
\tbl{Performance with the addition of demographic information, including models outlined in Section~\ref{sec:nonlinear}.}
{
\begin{tabular}{l l}
Model & AUPRC \\
\hline 
\hline
LR~\cite{rotmensch2017learning} & 0.1927\\
NB\cite{rotmensch2017learning} & 0.2515\\
\hline
LR (causal) & 0.1761\\
LR (causal, demo) & 0.1704\\
\hline
RF (causal) & 0.2493\\
RF (causal, demo) & 0.2509\\
\hline
NO\cite{rotmensch2017learning} & 0.3400\\
NO  (demo) & 0.3391\\
\end{tabular}
}
\label{tab:demo}
\end{minipage}
\begin{minipage}{3in}
\centering
\tbl{AUPRC of health knowledge graphs learned using noisy OR from data subsets.}
{
\begin{tabular}{l l r}
Model & AUPRC & Sample size \\
\hline
\hline
ED dataset & 0.3400 & 273,174 \\
\hline
Age: under 21 & 0.1971 & 15,467 \\
Age: 21-44 & 0.3039 & 97,510\\
Age: 45-64 & 0.3102 & 90,739\\
Age: 65-84 & 0.2391 & 51,550\\
Age: 85+ & 0.1653 & 17,909\\
\hline
Female & 0.3212 & 149,047\\
Male & 0.3053 & 124,128 \\
\\
\end{tabular}
}
\label{tab:conditional}
\end{minipage}
\end{table}

We build health knowledge graphs from noisy OR  parameter extraction for subsets of the patient population by age bracket and gender and report the AUPRC in Table ~\ref{tab:conditional}. We find that health knowledge graphs created from both genders have reasonable AUPRC while health knowledge graphs created from groups of smaller sample size or potentially many confounders perform poorly. Poor sample size can lead to model underfitting and poor performance. Additionally older patients may have more diseases (see Table~\ref{tab:cr_summary}) and additional health concerns that may make learning the relationships between diseases and symptoms harder.



We compare prior work~\cite{rotmensch2017learning} on an ED dataset to health knowledge graphs learned from a larger CR dataset from the same hospital. We present summary statistics broken down for the three forms of the CR dataset in Table~\ref{tab:cr_summary}. CR (single) refers to the dataset where each patient note corresponds to a separate visit. CR (episode) refers to a patient visit created by gaps smaller than 30 days long. CR (patient) refers to a patient's entire medical record over years denoting one patient visit. See Section~\ref{sec:data} for more details. For a distribution of diseases and symptoms by age group, we direct the reader to the supplementary materials.

\begin{table}[h]
    \centering
    \tbl{Median number of disease and symptom across patients by age group.}
    {
    \begin{tabular}{l r r r r r}
        Dataset & under 21 & 21-44 & 45-64 & 65-84 &  85+\\
        \hline 
        \hline
        Median symptoms (ED) & 2 & 2 & 2 & 2 & 2 \\
        Median symptoms (CR, single) & 0 & 0 & 1 & 1 & 1 \\
        Median symptoms (CR, episode) & 4 & 4 & 4 & 4 & 4 \\
        Median symptoms (CR, patient) & 7 & 14 & 29 & 37 & 29 \\
        \hline 
        Median diseases (ED) & 0 & 0 & 1 & 1 & 1 \\
        Median diseases (CR) & 0 & 0 & 1 & 1 & 1 \\
        Median diseases (CR, episode) & 1 & 2 & 2 & 3 & 3 \\
        Median diseases (CR, patient) & 2 & 5 & 12 & 17 & 15 \\
    \end{tabular}
    }
    \label{tab:cr_summary}
\end{table}

In Table~\ref{tab:cr}, we show the AUPRC for health knowledge graphs learned on the three CR datasets compared to the ED dataset. The CR datasets, as described further in Table~\ref{tab:cr_summary}, decrease in granularity with more symptoms and diseases observed for each patient visit. We observe that as granularity decreases, the noisy OR performance is comparable. Naive Bayes has a decreasing performance as the data becomes less granular, likely due to the violations in conditional independence that naive Bayes assumes. Logistic regression, however, appears able to discern more signal from the noise and increases in performance as granularity decreases. 

\begin{table}[h]
\centering
\tbl{AUPRC for ED and CR datasets.}
{
\begin{tabular}{l l l l l}
Model & ED dataset & CR (single) & CR (episode) &  CR (patient)\\
\hline
\hline
Logistic Regression & 0.1927 & 0.2332 & 0.2590 & 0.2678\\
Naive Bayes & 0.2515 & 0.2273 & 0.2129 & 0.1836\\
Noisy OR & 0.3400 & 0.3098 & 0.3147 & 0.3183\\
\end{tabular}
}
\label{tab:cr}
\end{table}

\section{Discussion}
\label{sec:discussion}

\subsection{Data size does not always matter.}

When evaluating models and individual diseases, we observe that sample size can have a large effect. For example, when evaluating on patient populations with different demographics, we find that demographics with smallest sample size correspond to poor AUPRC.

Importantly, however, drastically larger sample size does not necessarily guarantee drastically better performance, as evidenced with the lack of association between diseases and occurrence count. Additionally, the larger size of the online medical record dataset doesn't actually yield improvements in AUPRC. One hypothesis is that the emergency department is a key decision point in a patient's care where an exhaustive list of active symptoms and diseases are elicited and documented in order to diagnose and disposition a patient. However, given the limited resources of the emergency department, this process must be focused and is biased towards more acute complaints, so more chronic conditions and symptoms that are not contributory to the active problem may not be documented. 

The ED likely yields an adequate data set for most acute complaints, but likely inadequate for more chronic conditions. The CR dataset, on the other hand, also includes notes from providers who manage more chronic conditions, and therefore may be biased towards more chronic conditions and symptoms.

\subsection{Confounders may explain errors.}
Most strikingly, the number of co-occurring diseases has a high correlation with low F1 scores in our examination of the ED dataset. Put another way, the more diseases a patient has, the more error. We can imagine multiple reasons why. When a patient presents with multiple diseases, the corresponding symptoms may be difficult to attribute to the correct disease. Furthermore, one disease may in fact increase the susceptibility to other diseases and also the resulting symptoms. Lastly, multiple co-occurring diseases may be the result of an underlying problem that could affect diseases and symptoms. 

When we expand to the CR dataset, we increase the number of observed diseases and symptoms per instance from CR (single) to CR (episode) to CR (patient). The information gain in each visit may overwhelm models like the naive Bayes, which decreases in performance, while models like logistic regression seem to improve off the higher density of information. 

Another potential confounder is the selection of diseases and symptoms themselves in the evaluation GHKG. The original researchers~\cite{rotmensch2017learning} selected diseases and symptoms relevant to the acute emergency department for evaluation. Examining the broader CR datasets introduces the possibility that observed symptoms may be caused by diseases that are not in the scope of selection. One potential method to assess this confounding would be to widen the extraction process and see how the performance errors differs.

When considering extraction of medical knowledge, it is therefore important to consider the setting of application, the source of the dataset, and any potential differences. Controlled examination across multiple but similar data sources can illuminate potential confounders and suggest solutions.







\subsection{Increased model complexity does not necessarily help.}

We found that the inclusion of non-linear models does not necessarily improve AUPRC. The continued success of noisy OR models may be attributed to the high likelihood of false negatives and low likelihood of false positives in the dataset: an observed disease or symptom is likely true whereas an unobserved disease or symptom is possibly not true. 

While the non-linear models can take advantage of increased computational power and more expressive model parameterization, it does not appear that the strengths of non-linear models are able to compensate for the weaknesses of noisy OR. When leveraging advances in machine learning methods, it is important to understand potential sources of error to improve deficiencies in the existing models. 

\subsection{Limitations remain as an opportunity for future work.}

Our paper focuses on evaluation through a manual and curated GHKG for the emergency department setting. This focus affects all aspects of the pipeline, including which symptoms and diseases to evaluate. AUCPRC as reported is an imperfect measure of performance. Multiple evaluation methods could reduce additional bias.

In order for a health knowledge graph to truly aid clinicians and provide candidates of diseases, a health knowledge graph must also be assessed in the relevant clinical context. We have considered the emergency department, but other potential settings including primary care or before a patient enters a hospital would introduce an opportunity to more robustly assess model generalization. 

We grapple with confounders and questions of causal inference from observational data. Unmeasured confounders make determining causal relations difficult without additional experiments. Further work could include methods on improving causal inference methods or incorporating suggestions from our health knowledge graph into the clinical workflow.

Lastly, our model assumes a bipartite graph of binary edges between diseases and symptoms. Relaxing these constraints --- for example allowing one disease to cause another disease or allowing for different edge strengths --- may make the health knowledge graph more closely model medical knowledge and therefore clinically useful.

\section{Conclusion}
\label{sec:conclusion}

We assess the robustness of a health knowledge graph from patient visits from only the emergency department and the complete online medical record. We find that disease-specific performance varies widely and is correlated to factors like sample size and number of co-occurring diseases while heterogeneity of patients across age and gender causes differences in performance. Non-linear functions, although based on causal inference principles, do not lead to performance improvements. Lastly, we assess the generalizability on three datasets derived from a complete medical record for the same patients to better understand our models.

Researchers seeking to build health knowledge graphs might struggle with dataset questions including sufficient size and appropriate source. They may also be curious about algorithmic choices include whether a more simplistic model like logistic regression is adequate compared to more complex models. We describe several data factors of consideration including size, features, and sources of error. We show that the noisy OR approach is resiliently dominant in many problem settings. 

We believe the most promising areas of future research involve incorporating additional datasets across disparate medical settings in order to builds clinician trust as well as expose area for potential improvement. Health knowledge graphs can be used to provide automated recommendations for clinicians as well as find new edges between diseases and symptoms to advance understanding of disease.

\section*{Acknowledgements}

The authors thank Maggie Makar and Rebecca Boiarsky for helpful comments and Google, Inc. for permitting the use of the Google health knowledge graph within our evaluation. This work is supported by a grant from Google, Inc.

\bibliographystyle{ws-procs11x85}
\bibliography{bibliography}

\appendix

\section{Datasets}
\label{sec:datasets}

\subsection{Demographic information}
For models that allow for continuous features, e.g. Logistic Regression, we augment the existing disease and symptom observations with the age and binarized sex values. For models that only allow binary features, we create age bracket (less than 21, 21-44, 45-64, 65-84, 85+) and include a binary indicator for each age bracket if the patient’s age is within that bracket. For sex, we include one binary feature if the patient is female and one binary feature if the patient is male. 

\subsection{Data distributions}

To better understand the datasets used, we present the distribution of diseases and symptoms for the dataset of emergency department (ED) patient visits and the three complete record (CR) datasets.

\begin{figure}
    \centering
    \includegraphics[width=0.4\textwidth]{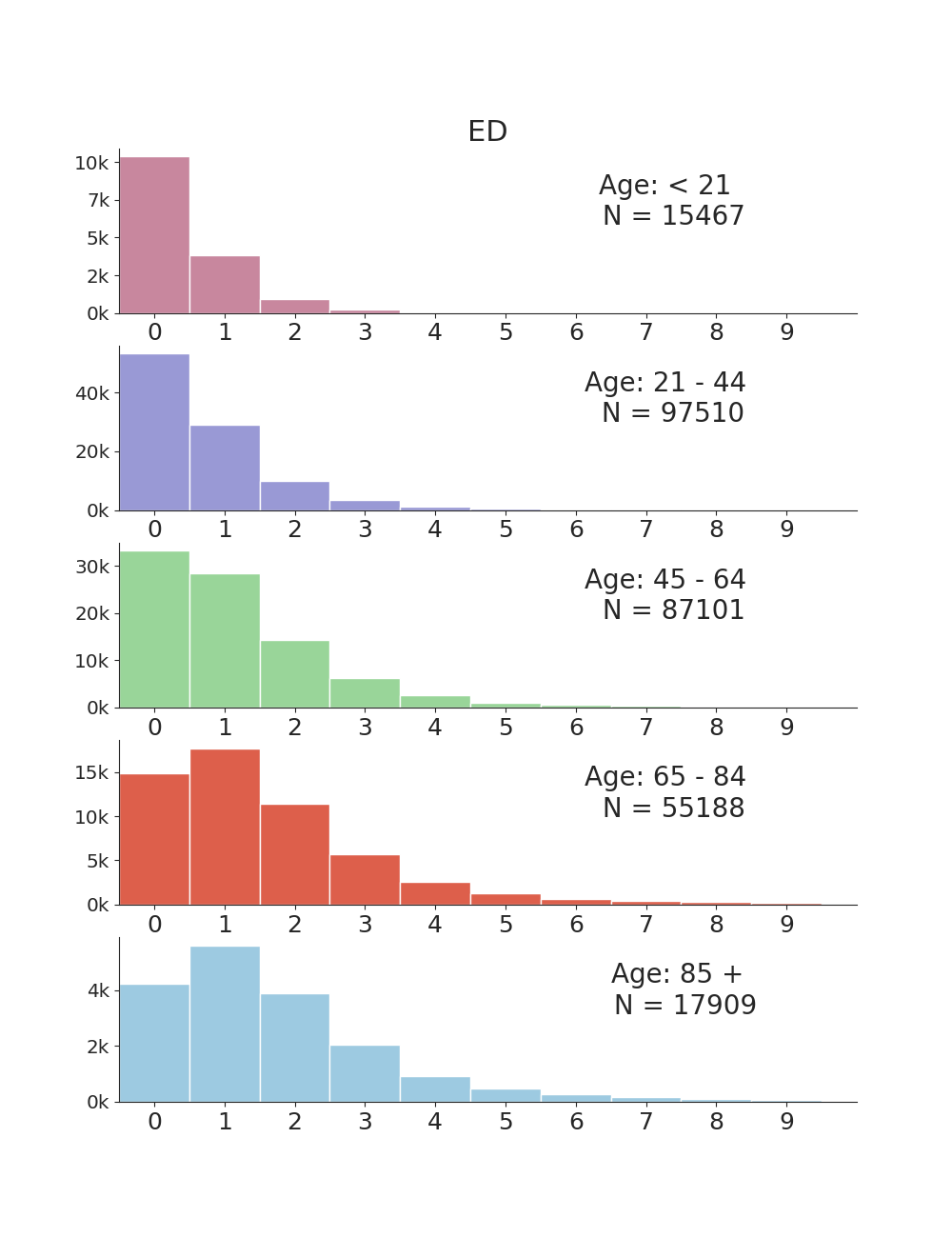}
    \includegraphics[width=0.4\textwidth]{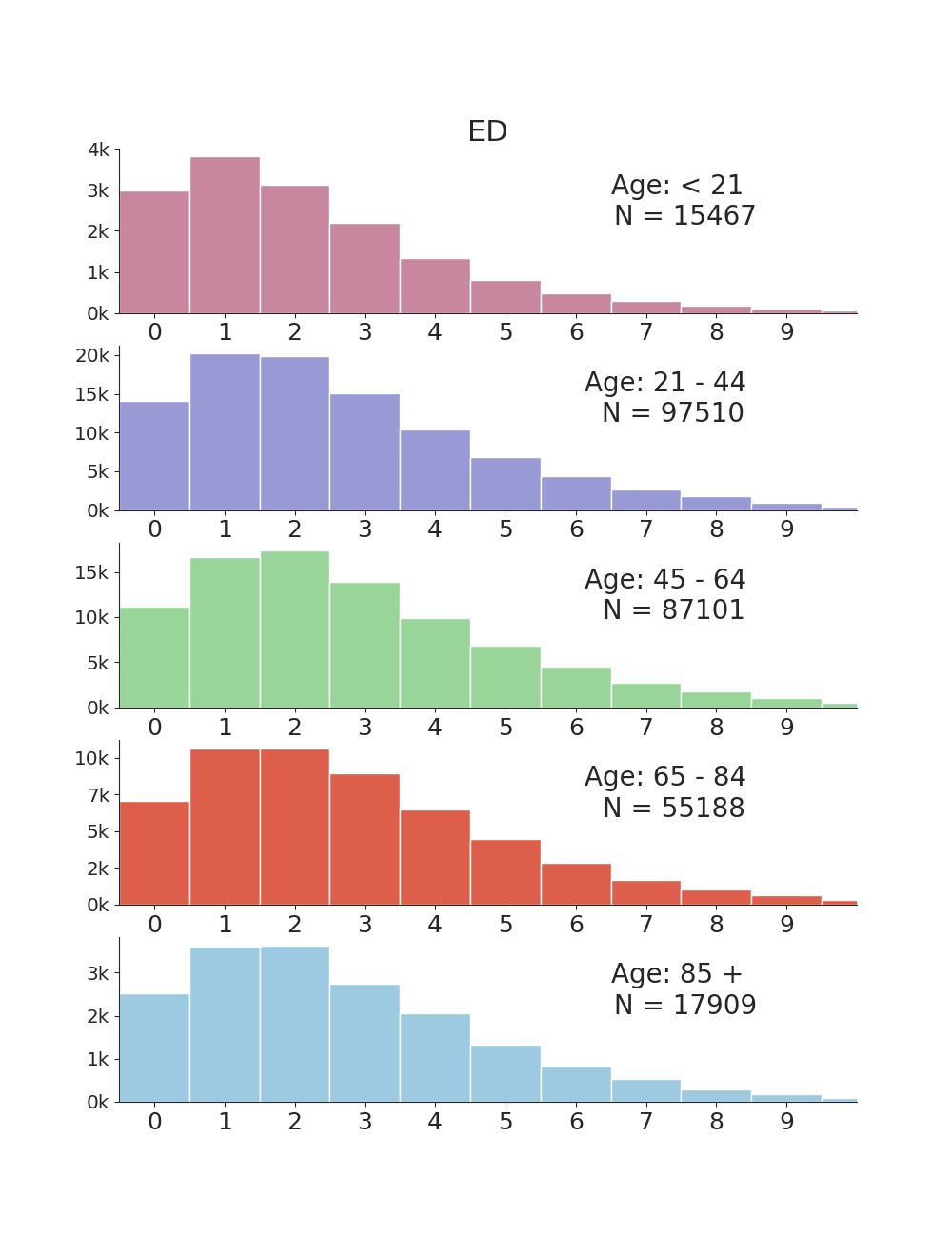}
    \caption{Distribution of diseases (left) and symptoms (right) for emergency department dataset (ED).}
    \label{fig:ed}
\end{figure}

\begin{figure}
    \centering
    \includegraphics[width=0.4\textwidth]{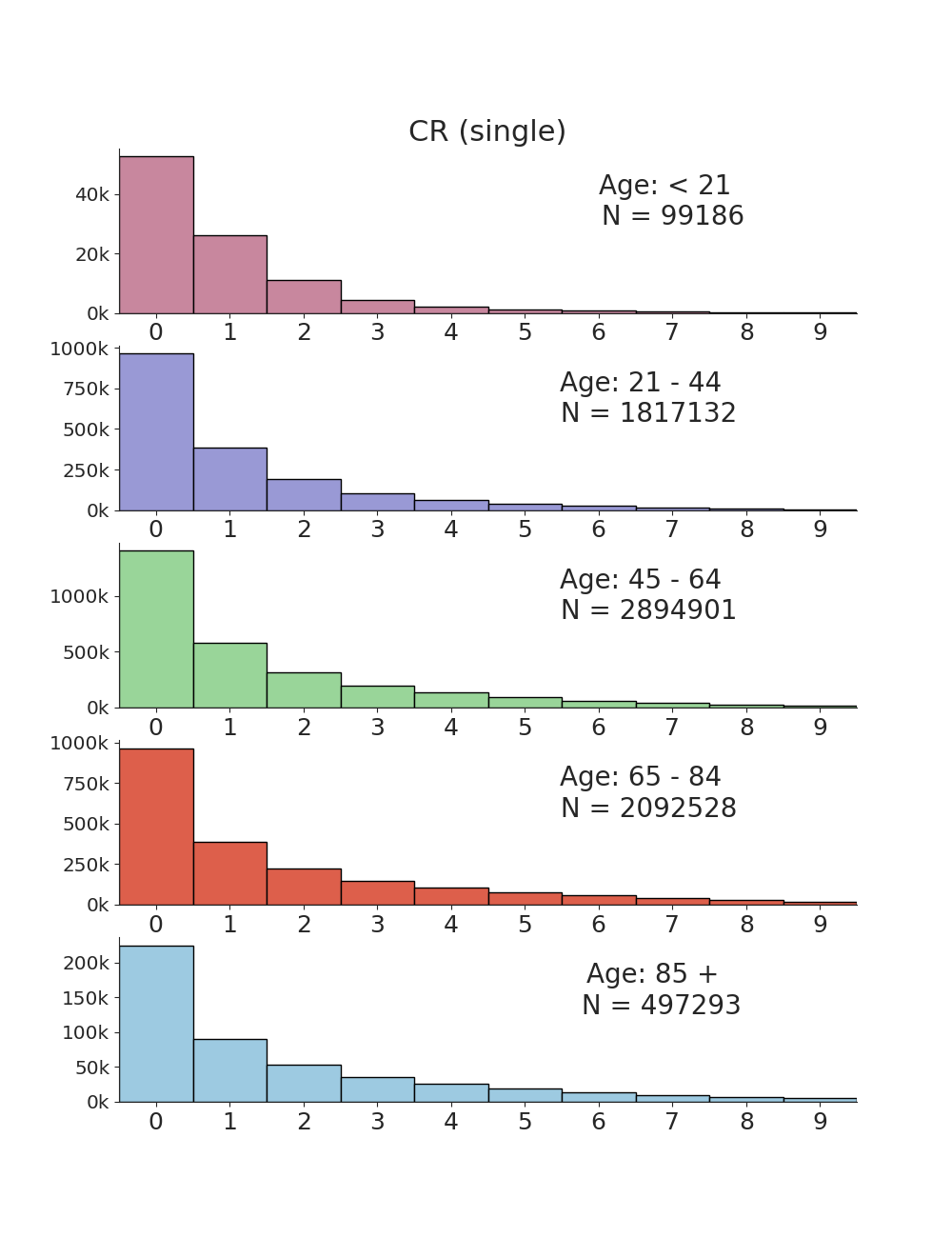}
    \includegraphics[width=0.4\textwidth]{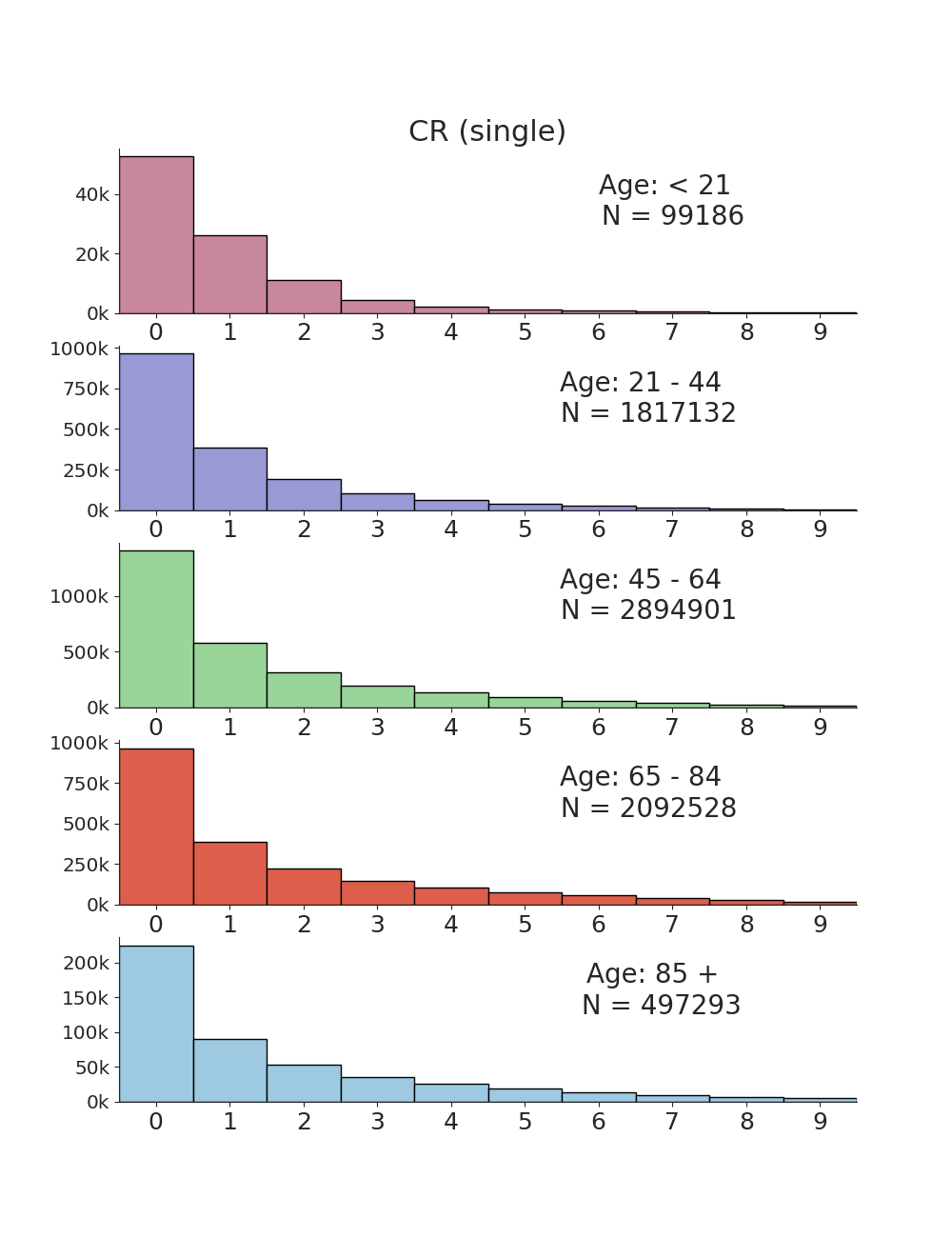}
    \caption{Distribution of diseases (left) and symptoms (right) for complete record dataset split into single patient visits (CR, single).}
    \label{fig:cr_single}
\end{figure}

\begin{figure}
    \centering
    \includegraphics[width=0.4\textwidth]{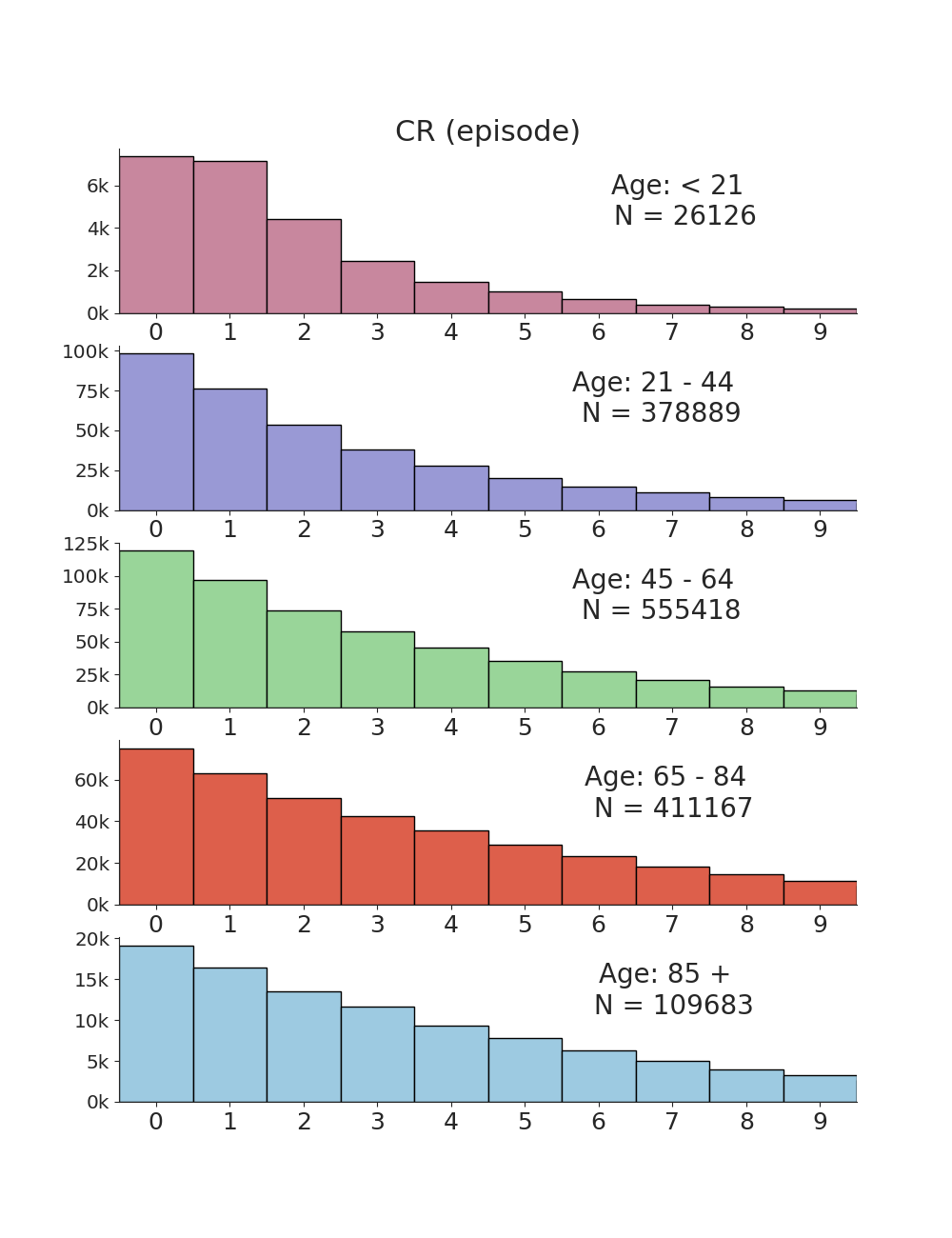}
    \includegraphics[width=0.4\textwidth]{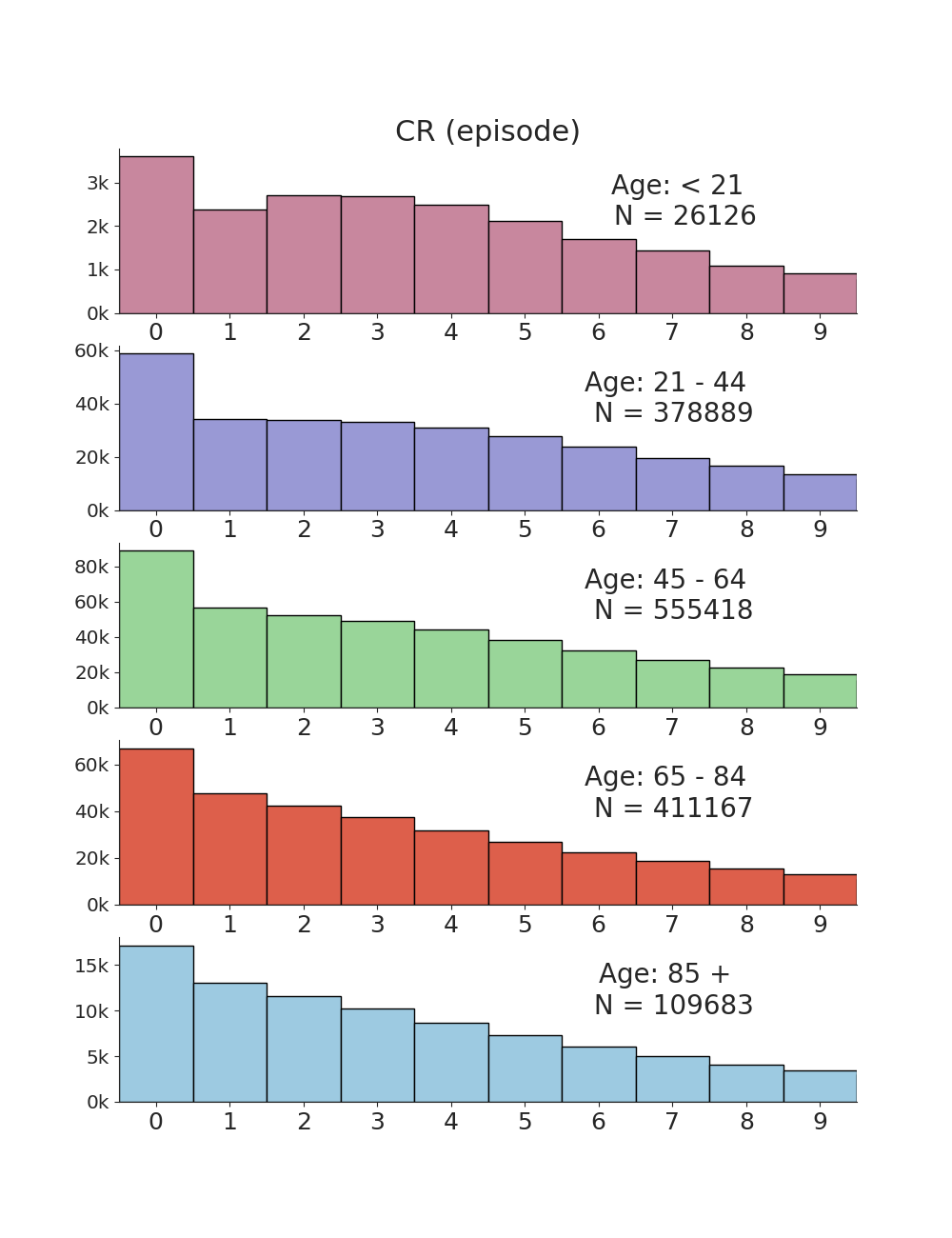}
    \caption{Distribution of diseases (left) and symptoms (right) for complete record dataset split into episodes (CR, episode).}
    \label{fig:cr_episode}
\end{figure}

\begin{figure}
    \centering
    \includegraphics[width=0.4\textwidth]{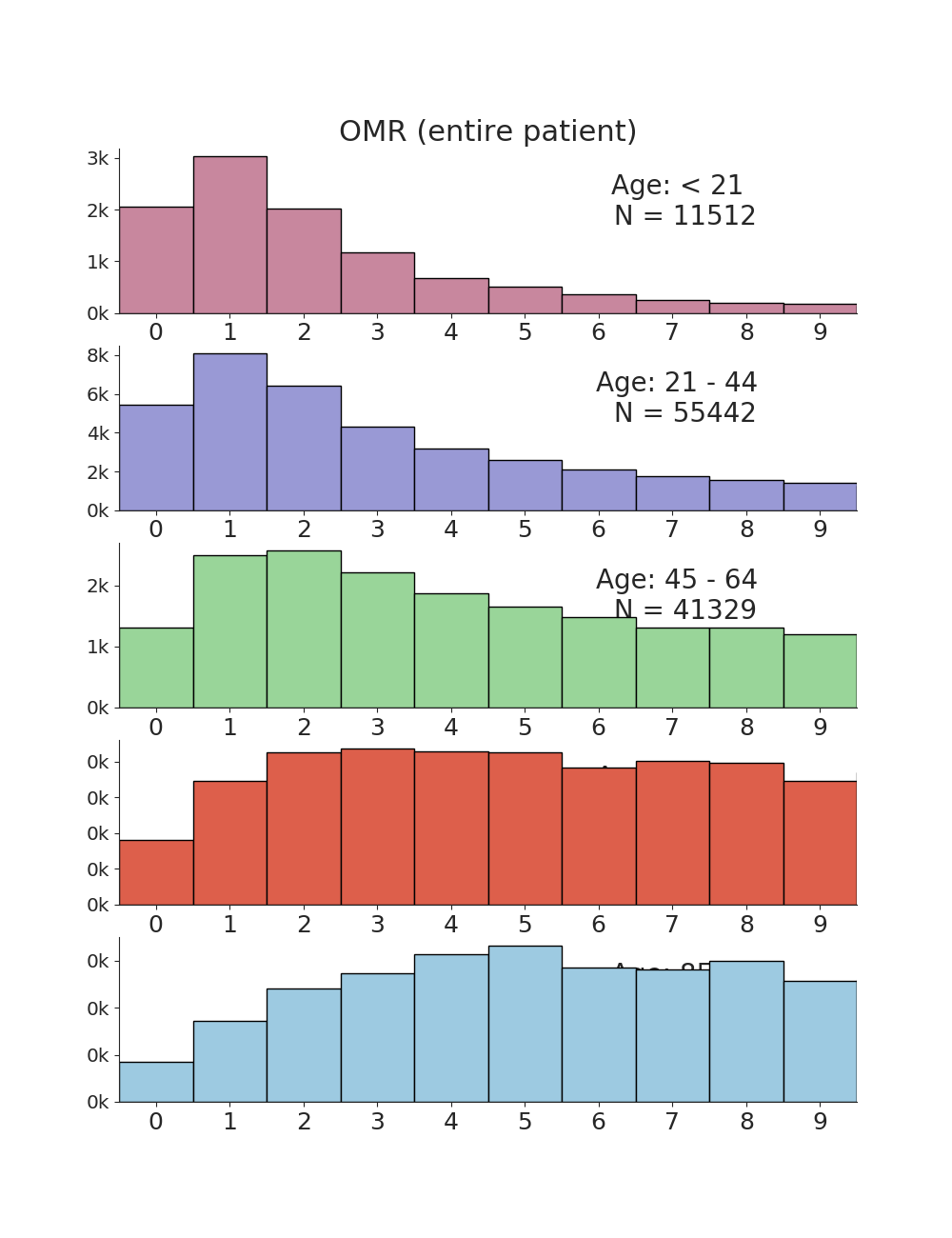}
    \includegraphics[width=0.4\textwidth]{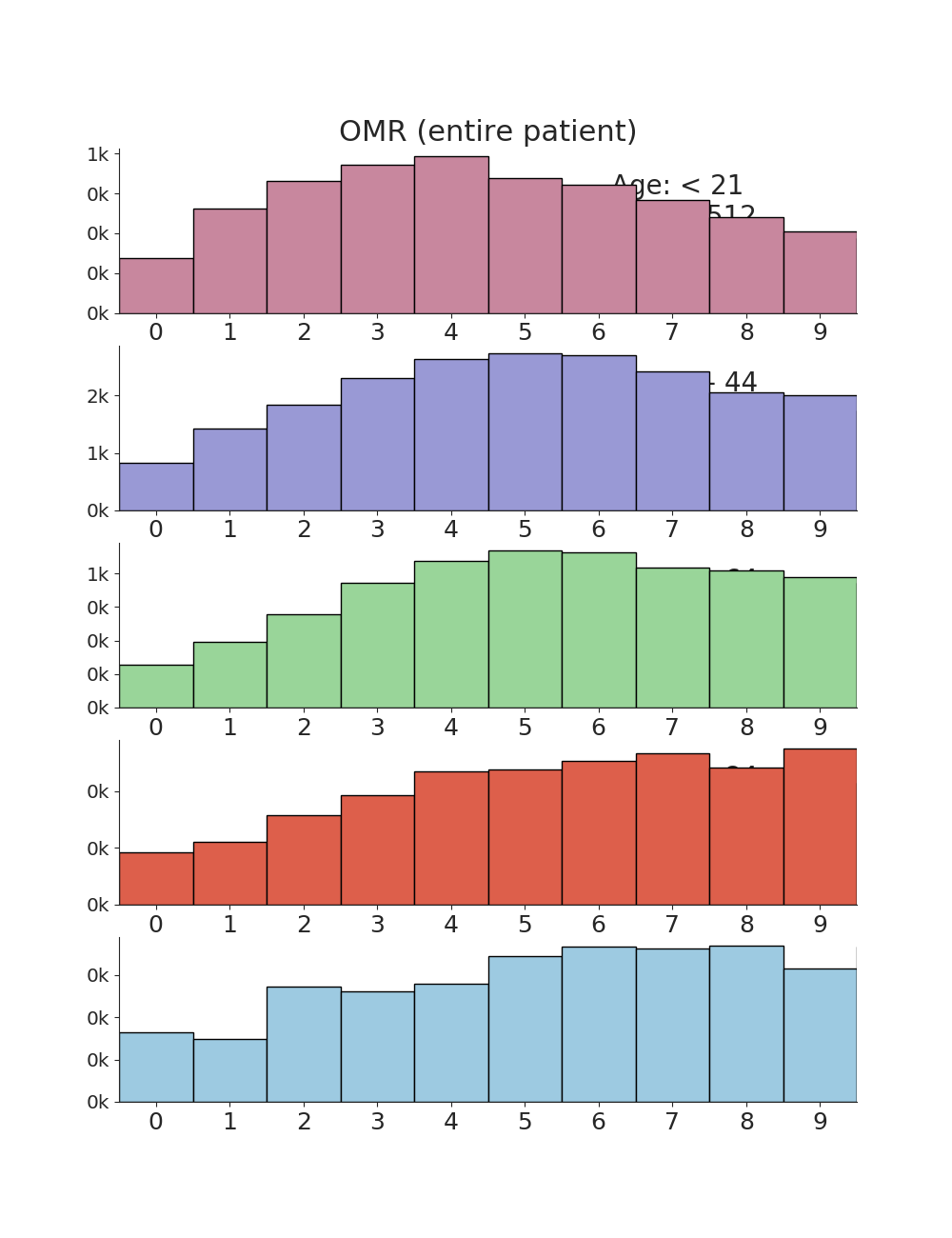}
    \caption{Distribution of diseases (left) and symptoms (right) for complete record dataset split across entire patient histories (CR, patient).}
    \label{fig:cr_patient}
\end{figure}

\begin{figure}
    \centering
    \includegraphics[width=0.4\textwidth]{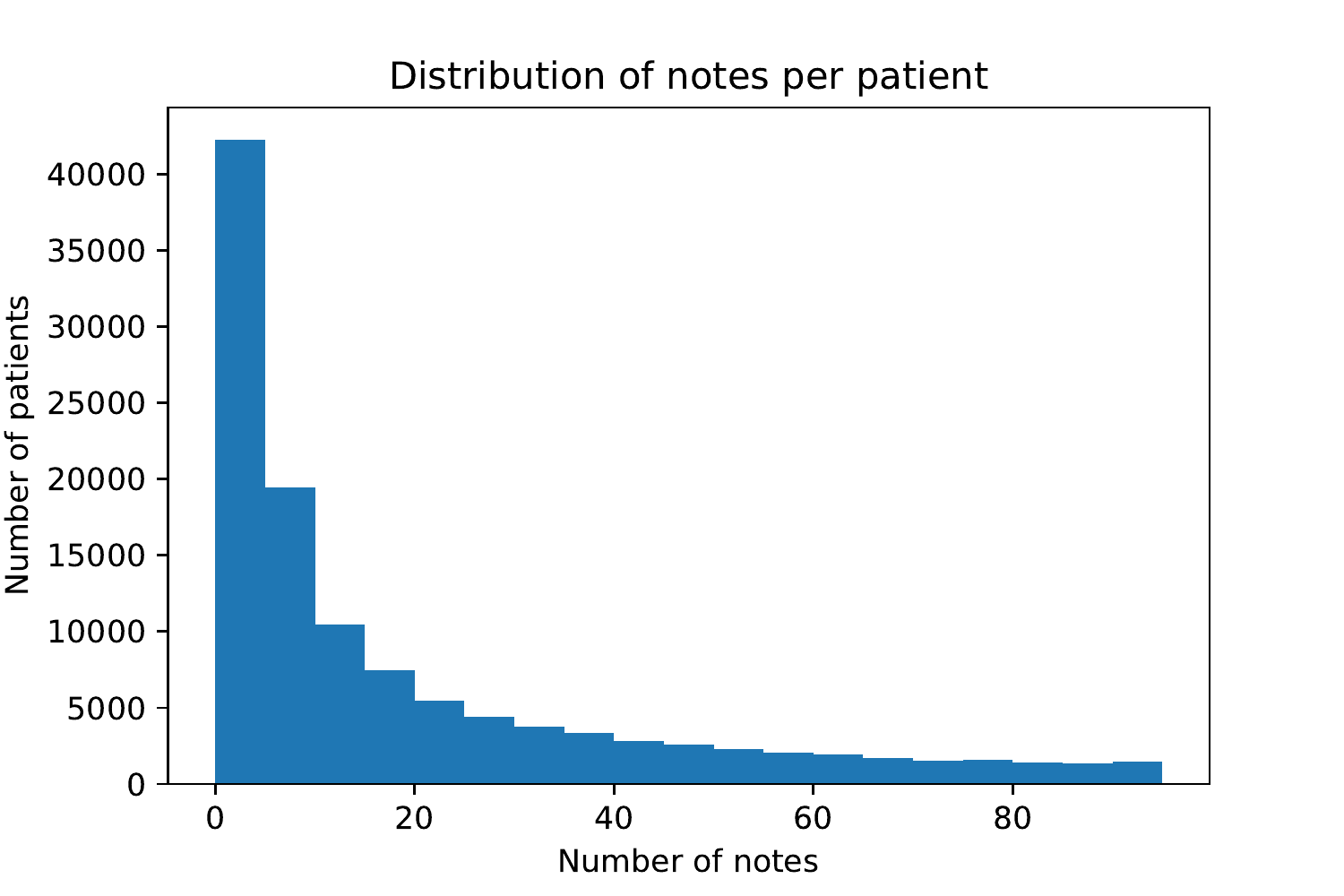}
    \includegraphics[width=0.4\textwidth]{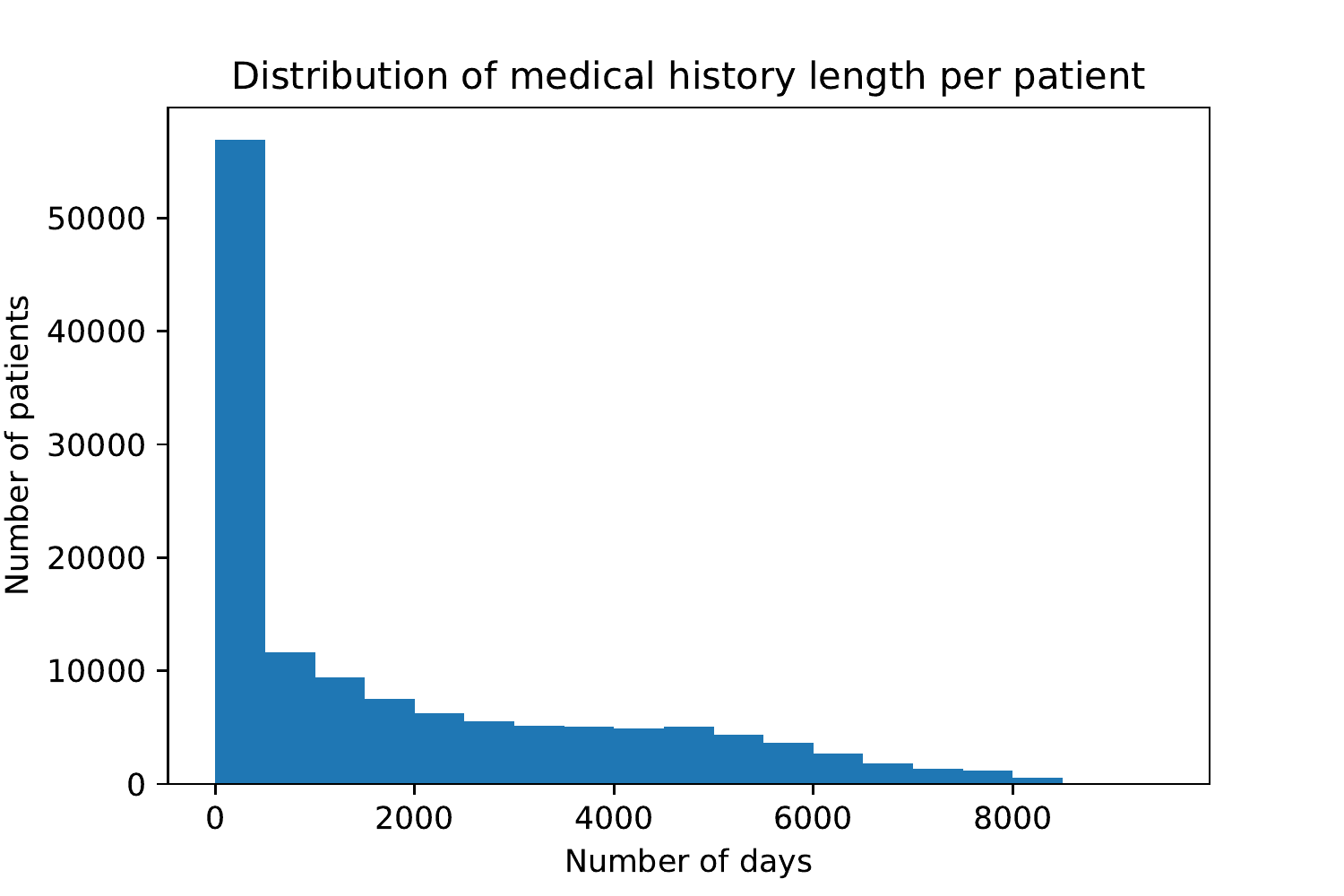}
    \caption{Distribution of the number of notes per patient in the CR dataset (left), and the distribution of the time spanned in days for each patient by their CR record (right).}
    \label{fig:cr_overall}
\end{figure}

\begin{figure}
    \centering
    \includegraphics[width=0.5\textwidth]{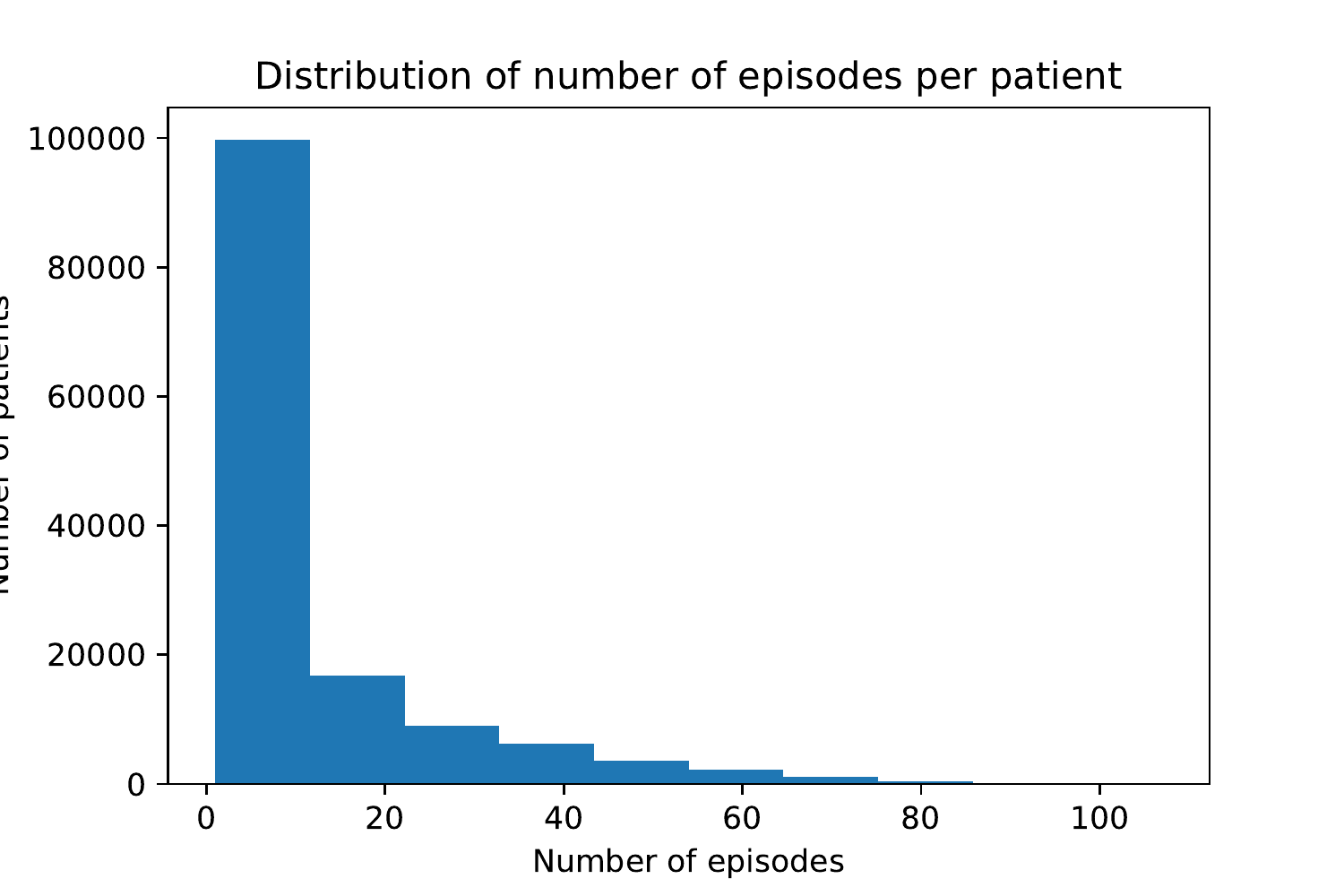}
    \caption{Distribution of the number of episodes per patient in the CR dataset.}
    \label{fig:num_episodes}
\end{figure}

\begin{figure}
    \centering
    \includegraphics[width=0.4\textwidth]{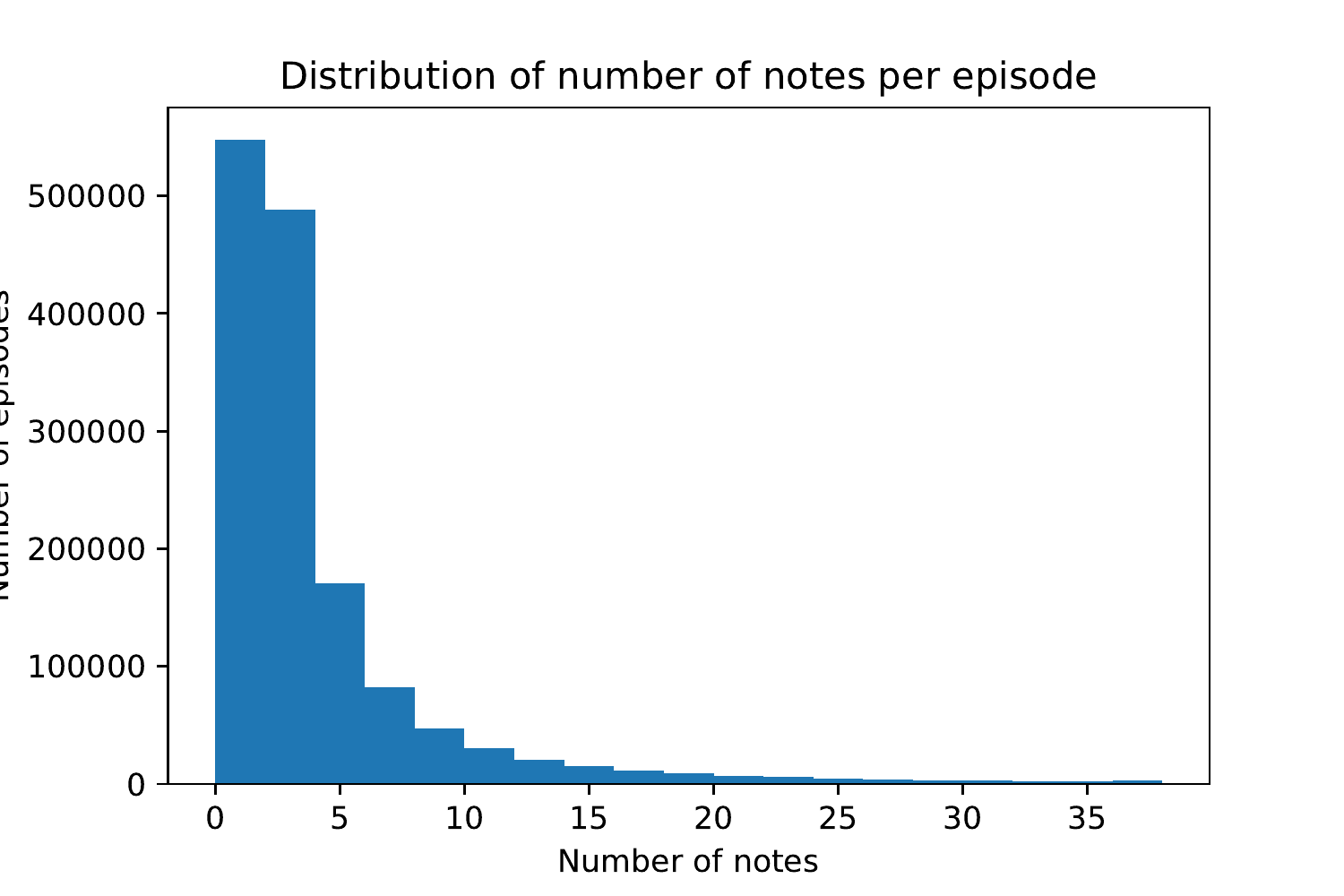}
    \includegraphics[width=0.4\textwidth]{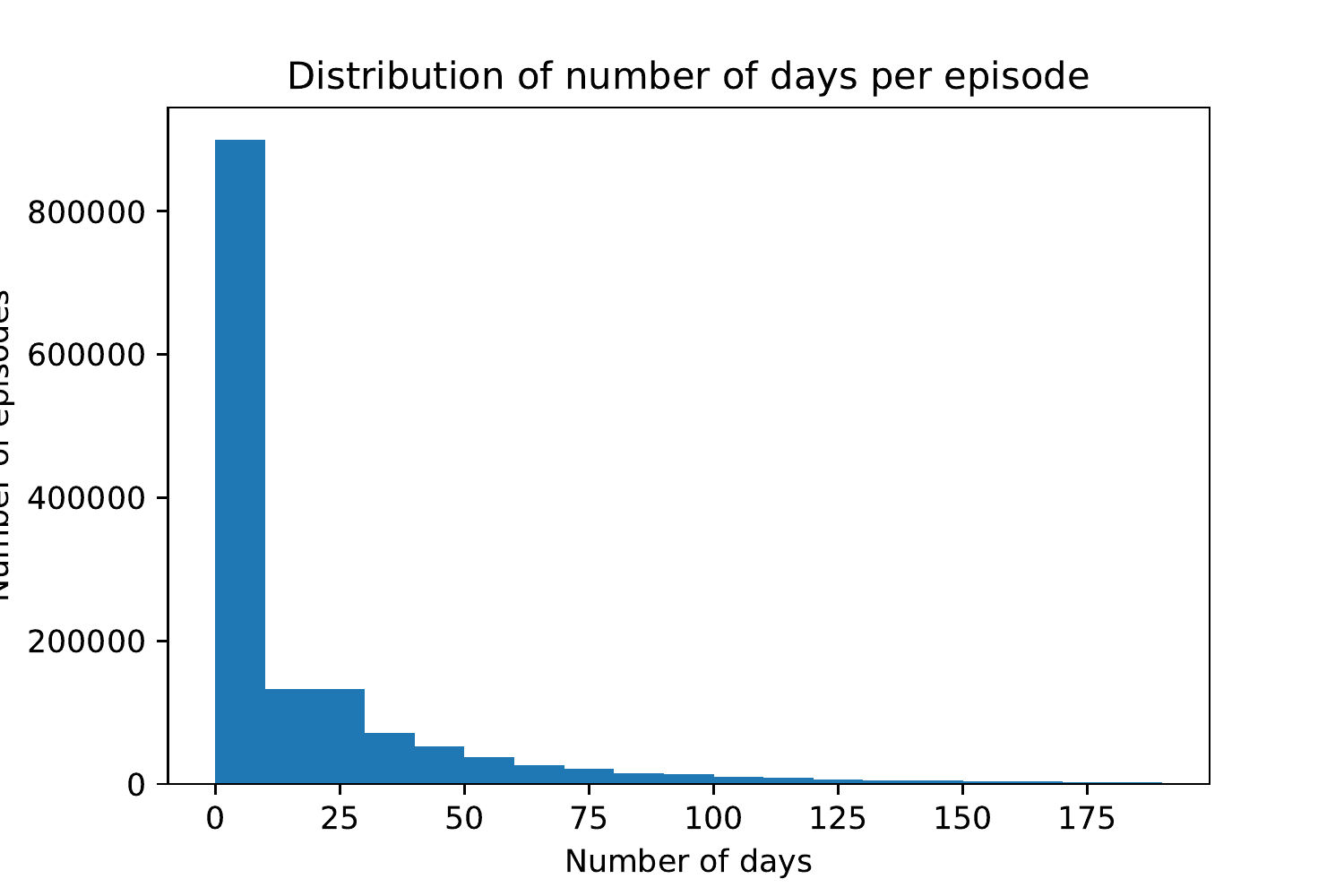}
    \caption{The distribution of lengths of each episode in number of notes (left) and number of days (right).}
    \label{fig:cr_lengths}
\end{figure}

Next, to better understand the CR dataset, we examined the distributions of notes and episodes in Figure ~\ref{fig:cr_overall} -  Figure~\ref{fig:cr_lengths}. Figure ~\ref{fig:cr_overall} shows that there is a long tail in both the number of notes per patient, and the time span of notes available for each patient. In Figure ~\ref{fig:num_episodes}, we see that the same is true for the number of created episodes per patient. Figure ~\ref{fig:cr_lengths} shows that the vast majority of episodes are fewer than 5 days and 5 notes long.

\section{Evaluation metrics}

The GHKG contains the binary disease-symptom pairs for diseases that cause symptoms whereas our models for learning health knowledge graphs from EHRs return an importance metric $\delta_{ij}$ for symptom $i$ and disease $j$. 

For individual disease analysis, we use the F1 score. For a given disease, find the top $E_j$ importance scores $\delta_{ij}$ where $E_j$ is the number of symptoms in the GHKG for disease $j$. For those symptoms, designate those disease-symptom pairs as selected. Then compute the F1 score $F_1$ according to $F_1 = \frac{2TP}{2TP + FP + FN}$ where: 1) $TP$ is the number of true positives or disease-symptom edges that are selected by our model and also the GHKG, 2) $FP$ is the number of false positives or disease-symptom edges that are selected by our model but not selected by the GHKG, and 3) $FN$ is the number of false negatives or disease-symptom edges that are not selected by our model but are selected by the GHKG.

\begin{figure}
    \centering
    \includegraphics[width=0.8\textwidth]{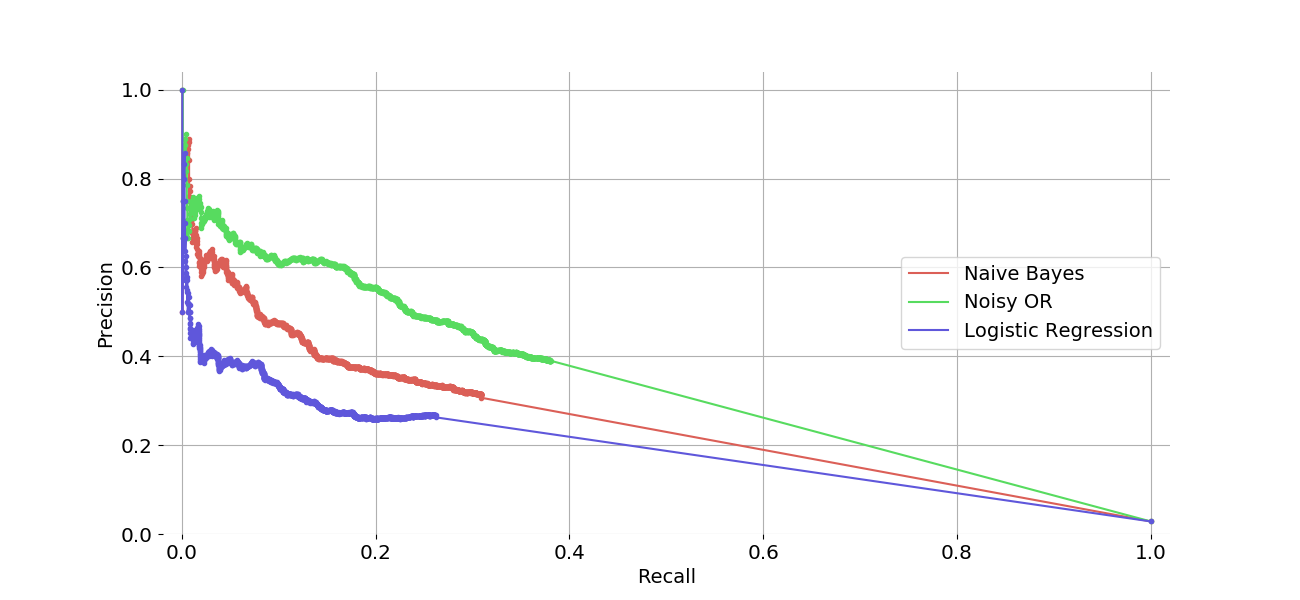}
    \caption{Example AUPRC curves with added point to extend curves to recall=1. Area found with trapezoidal approximation.}
    \label{fig:auprc}
\end{figure}

The area under the precision-recall curve (AUPRC) is computed as follows. For a given disease, find the top $E_j$ importance scores $\delta_{ij}$ where $E_j$ is the number of symptoms in the GHKG for disease $j$. For those symptoms, include those importance scores $\delta_{ij}$ in the precision-recall computation and 0 otherwise. For all non-zero disease-symptom scores, compute the precision and recall according to precision = $\frac{TP}{TP+FP}$ and recall = $\frac{TP}{TP+FN}$ with differing thresholds to compute different coordinates of recall and precision. Because precision-recall curves may end at different locations,\cite{flach2015precision} we extend each curve to the point $(1,B)$ where $B$ corresponds to the precision value when recall is set to 1. In order words, when we select every edge, what is the resulting precision? $B$ then becomes the fraction of selected edges in the GHKG compared to the total possible edges. See Figure~\ref{fig:auprc} for an illustration of example AUPRC curves and the extension to $(1,B)$. The area under the curve is then found with a trapezoidal approximation.

\section{Error Analysis}
\label{sec:error_analysis}
Here we give a more precise description of the disease error analysis performed. 

\begin{itemize}
    \item For every disease, count the number of patient visits where this disease was observed. We denote this value the number of occurrences. A disease is considered abnormal if the number of occurrences is then a standard deviation below the mean number of occurrences across all diseases. (\texttt{count})
    \item For every disease, find all patient visits where this disease was observed. For each of these patient visits, total the number of observed diseases in each patient visit. For each disease, compute the mean number of diseases over all patient visits. We denote this value the mean number of extracted diseases. A disease is considered abnormal if mean number of extracted diseases is a standard deviation above the mean number of extracted diseases across all diseases. (\texttt{disease})
    \item For every disease, find all patient visits where this disease was observed. For each of these patient visits, total the number of observed symptoms in each patient visit. For each disease, compute the mean number of symptoms over all patient visits. We denote this value the mean number of extracted symptoms. A disease is considered abnormal if mean number of extracted symptoms is a standard deviation above the mean number of extracted symptoms across all diseases. (\texttt{symptom})
    \item For every disease, find all patient visits where this disease was observed. For each of these patient visits, extract the patient age at the time of visit. For each disease, compute the mean age over all patient visits. We denote this value the mean patient age. A disease is considered abnormal if mean patient age a standard deviation either above or below the population mean. (\texttt{age})
    \item For every disease, find all patient visits where this disease was observed. For each of these patient visits, extract the patient gender. For each disease, compute the percentage female over all patient visits. We denote this value the female percentage. Because the clinical records require all patients to be either male or female, we can find the inverse by subtracting female percentage from 1. A disease is considered abnormal if female percentage is a standard deviation either above or below the population female percentage. (\texttt{female})
    \item any of the above abnormalities (\texttt{any})
\end{itemize}

\section{Disease predictability}

We are interested in the predictive ability of our models to predict the disease from the symptoms. Although our main evaluation metrics use the GHKG, which has been manually curated by experts, the disease predictability can potentially shed light on whether we have extracted the correct symptoms for each disease. 

\begin{figure}
    \centering
    \includegraphics[width=0.8\textwidth]{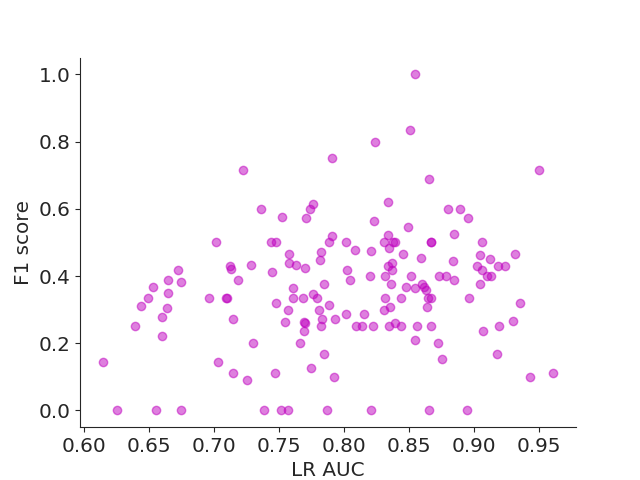}
    \caption{Comparison of AUC and F1 score for logistic regression.}
    \label{fig:auc_f1}
\end{figure}

\begin{figure}
    \centering
    \includegraphics[width=0.8\textwidth]{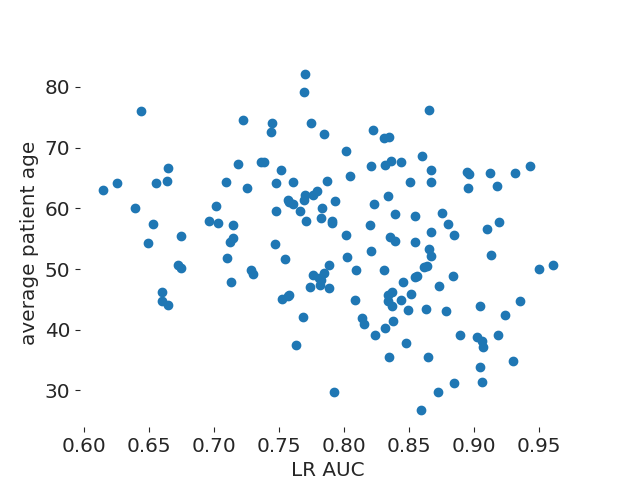}
    \caption{Relationship between patient average age and logistic regression AUC.}
    \label{fig:lr_avg_age}
\end{figure}

We compute disease predictability using a 3-fold cross-validated area under the received operator curve (AUC)~\cite{fawcett2006introduction}, searching over the same parameters as outlined in Section~\ref{sec:methods}. We report the average AUC using the best parameters. In Figure~\ref{fig:auc_f1}, we see that the relationship between logistic regression AUC and F1 score is not very strong. Additionally, in figure~\ref{fig:lr_avg_age} we see a negative relationship between logistic regression AUC and average patient age. This finding points to investigating younger patients and exploring why predicting for them is more difficult than older patients. One hypothesis is that younger patients don't manifest symptoms as evidently as older patients --- or that our extraction process omits symptoms that affect younger patients more.

\section{Symptom predictability and causal method performance}
\begin{figure}[h]
    \centering
    \includegraphics[width=0.42\textwidth]{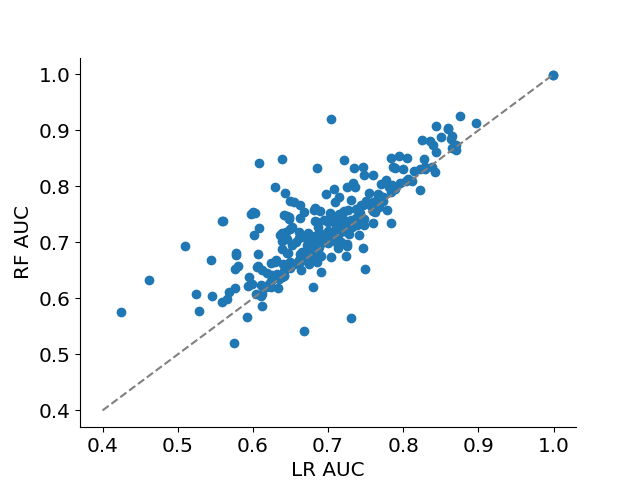}
    \includegraphics[width=0.42\textwidth]{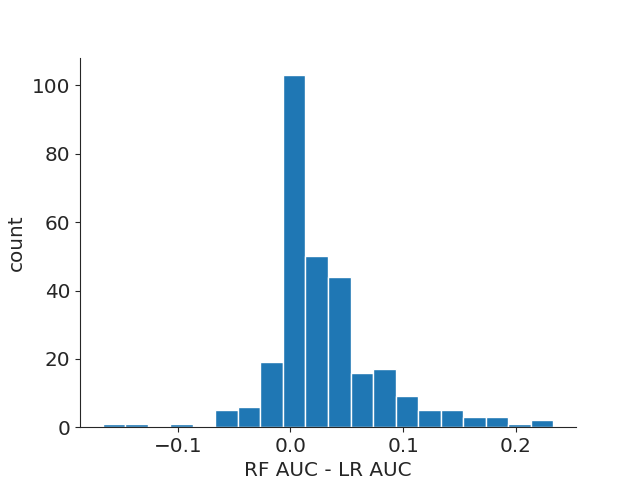}
    \caption{\textbf{Left:} Relationship between logistic regression and random forest AUCs for the same symptom. \textbf{Right:} Histogram of differences between logistic regression and random forest AUC for the same symptom.}
    \label{fig:sympt_lr_rf}
\end{figure}

In order to understand better the non-linear methods in Section~\ref{sec:nonlinear}, we can also investigate the predictability of symptoms. That is, for observed diseases, how well can we predict the symptoms? Across the different link functions used, how much variance in symptom AUC is there? Similar to disease predictability, we use 3-fold cross validation to determine the symptom AUC. We report the average AUC found using the optimal parameters. In figure~\ref{fig:sympt_lr_rf} we see that logistic regression and random forest AUC are correlated but can differ. Depending on the sample size and linearity of the underlying data generating function, different models may be better suited for different symptoms. One area for future research might be then to learn which predictive model to use for each symptom independently in order to build a more accurate health knowledge graph.

Additionally, we investigate if the random forest causal method ever outperforms the noisy OR method. Over the 156 possible diseases, we identify one disease where the F1 score for the random forest causal method outperforms the noisy OR method: breast cancer. For the other diseases, either the F1 scores match or the noisy OR method performs better.

\section{CR dataset disease analysis}

Here we present addition results on the CR datasets, specifically the disease analysis outlined in Section~\ref{sec:error_analysis}. 

\begin{table}[h]
    \centering
    \tbl{Percentage of diseases with abnormalities learned on CR (single) data.}
    {
    \begin{tabular}{l  c c c c c  c}
         & \texttt{count} & \texttt{disease} & \texttt{symptom} & \texttt{age} & \texttt{female} & \texttt{any}  \\
        \hline
        \hline
top 50 & 16\% & 12\% & 0\% & 10\% & 4\% & 38\% \\
bottom 50\%& 14\% & 38\% & 6\% & 6\% & 16\% & 60\% \\

    \end{tabular}
    
    \label{tab:top_bottom_single}
    }
\end{table}

\begin{table}[h]
    \centering
    \tbl{Percentage of diseases with abnormalities learned on CR (episode) data.}
    {
    \begin{tabular}{l  c c c c c  c}
         & \texttt{count} & \texttt{disease} & \texttt{symptom} & \texttt{age} & \texttt{female} & \texttt{any}  \\
        \hline
        \hline
top 50 & 12\% & 14\% & 0\% & 10\% & 10\%& 32\% \\
bottom 50 & 18\% & 40\% & 4\% & 14\% & 16\% & 57\% \\

    \end{tabular}
    
    \label{tab:top_bottom_episode}
    }
\end{table}

\begin{table}[h]
    \centering
    \tbl{Percentage of diseases with abnormalities learned on CR (patient) data.}
    {
    \begin{tabular}{l c c c c c c}
         & \texttt{count} & \texttt{disease} & \texttt{symptom} & \texttt{age} & \texttt{female} & \texttt{any}  \\
        \hline
        \hline 
        top 50 & 10\% & 14\% & 2\% & 14\% & 10\% & 36\% \\
        bottom 50 & 20\% & 32\% & 4\% & 20\% & 18\% & 64\% \\
    \end{tabular}
    
    \label{tab:top_bottom_patient}
    }
\end{table}

\end{document}